\documentclass[%
 reprint,
superscriptaddress,
 amsmath,amssymb,
 aps,
]{revtex4-2}

 \usepackage{hyperref}
 \hypersetup{
    colorlinks=true,       
    linkcolor=blue,          
    citecolor=blue,        
    filecolor=magenta,      
    urlcolor=blue           
}

\usepackage{graphicx}
\usepackage{xcolor}
\usepackage{dcolumn}
\usepackage{bm}
\usepackage{appendix}

\usepackage{mathtools,rotating} 

\usepackage[version=4]{mhchem}
\usepackage{booktabs}

\begin{document}

\preprint{APS/123-QED}

\title{Many-body theory calculations of positron binding to parabenzoquinone}
\author{S. K. Gregg} \email{sgregg07@qub.ac.uk}
\affiliation{
Centre for Light-Matter Interactions, School of Mathematics and Physics, Queen's University Belfast, University Road, Belfast BT7 1NN, United Kingdom.
}
 \author{J. Hofierka}
 \affiliation{
Theoretische Chemie, Physikalisch-Chemisches Institut, Universität Heidelberg, Heidelberg D-69120, Germany.
}
 \author{B. Cunningham}
 \affiliation{
Centre for Light-Matter Interactions, School of Mathematics and Physics, Queen's University Belfast, University Road, Belfast BT7 1NN, United Kingdom.
}


\author{D. G. Green} \email{d.green@qub.ac.uk}
\affiliation{
Centre for Light-Matter Interactions, School of Mathematics and Physics, Queen's University Belfast, University Road, Belfast BT7 1NN, United Kingdom.
}

\date{\today}

\begin{abstract}
Positron binding in parabenzoquinone is studied using \textit{ab initio} many-body theory. 
 The effects of electron-positron correlations including polarization, virtual positronium formation and positron-hole repulsion, as well as those of
 $\pi$ bonds, aromaticity, and lone electron pairs, are considered.
The binding energy is calculated as 60$\pm$16 meV, considerably larger than the 0.0925 meV value inferred from recent scattering calculations of [G. Moreira and M. Bettega, {\emph{Eur.~Phys.~J.~D}} {\bf 78} (2024)], but substantially smaller than we find in benzene (148$\pm$26 meV).
The positron contact density (lifetime) is calculated as 8.0$\times10^{-3}$ a.u. (2.48 ns), vs.~1.61$\times 10^{-2}$ a.u. (0.81 ns) in benzene. 
The decrease (increase) in binding (annihilation rate) in parabenzoquinone compared to benzene is ascribed  to the loss of aromaticity: the electron density on the positive oxygen nuclei being relatively harder for the positron to probe compared to the aromatic rings in benzene.
\end{abstract}
                  
\maketitle

\section{\label{sec:introduction} Introduction}
Positron interactions with molecules are characterized by strong many-body correlations between the molecular electrons and the positron, which makes these interactions challenging to accurately model.
Positron binding has been measured for over 100 molecules \cite{Gilbert02,Danielson09,Danielson10,Danielson12,Danielson12a,ArthurBaidoo2024} and recent progress in the theory of positron-molecule interactions has lead to the prediction of positron binding to a wide range of molecules, including alkanes, polar molecules, amino acids and ringed molecules \cite{Tachikawa2011, Koyanagi2012, Charry2014, Swann2019, Suguira2020, Hofierka2022, Cassidy2023, Hofierka2024, ArthurBaidoo2024}. 
Our many-body theory method provides an accurate \textit{ab initio} approach to calculating positron-molecule binding energies and facilitates the systematic inclusion of different types of correlation effects. The approach remains in its infancy, but has demonstrated considerable success in predicting positron-molecule bound state energies in good agreement with experiment \cite{Hofierka2022, Cassidy2023, Hofierka2024, ArthurBaidoo2024}, and also has proven capability to describe positron-molecule scattering \cite{Rawlins2023} and positronic bonding \cite{Cassidy2024_2}.

\begin{figure}[h]
    \centering
    \includegraphics[width=0.6\linewidth]{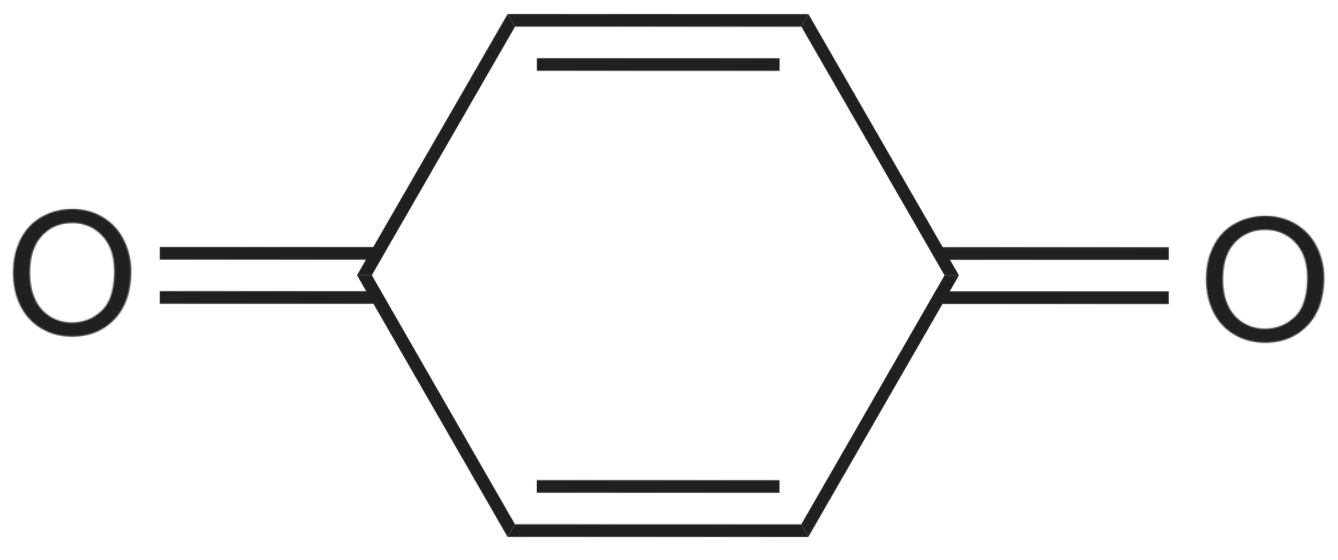}
    \caption{Structure diagram of the parabenzoquinone molecule. Single (double) lines represent single (double) bonds. Oxygen (O) atoms are labelled, and unlabelled vertices are carbon atoms with a number of hydrogen atoms attached such that each of the four valence electrons of the carbon atom is involved in a bond.
    }
    \label{fig:pbnzq_line_diagram}
\end{figure}

In the present paper, we focus on the parabenzoquinone (pBQ) molecule, C$_6$H$_4$O$_2$. Scattering calculations by Moreira and Bettega using the Schwinger multichannel method inferred the presence of a positron-pBQ bound state with binding energy 0.0925 meV \cite{Moreira2024}. Here, we use many-body theory to calculate the energy of the positron bound state in pBQ, the corresponding positron wavefunction and the contact density (positron lifetime). 
Parabenzoquinone is a planar ringed molecule with a six-carbon ring: four of the carbon atoms have hydrogen atoms bonded to them and the remaining two, which are at opposite sides of the ring, are bonded to oxygen atoms. 
The structure of this molecule is shown in Figure \ref{fig:pbnzq_line_diagram}. 
This molecule is similar in structure to benzene, for which positron binding has been previously studied using our many-body theory \cite{ArthurBaidoo2024}, leading to a natural comparison between the two molecules. Notably, for benzene our \emph{ab initio} approach highlighted quantitatively the enhanced importance of $\pi$ bonds on the strength of the positron-molecule correlation potential and thus binding energies \cite{Hofierka2022,ArthurBaidoo2024}, confirming previous speculation by experiment and model calculations \cite{Danielson09,Suguira2020}. In contrast to benzene, pBQ is not aromatic; whilst there are $\pi$-bonded molecular orbitals (MOs), the electron density is localised on the lone electron pairs (non-bonding MOs) on the oxygen atoms. Here, we find that compared to benzene, the lack of aromaticity in pBQ leads to a reduced binding energy.

The contents of this paper are as follows. Section \ref{sec:theory} briefly outlines the many-body theory used in the binding calculation. Section \ref{sec:results} contains our calculated binding energies and positron wavefunction, discusses the effect of $\pi$ bonds and lone electron pairs on the positron bound state and compares the positron bound state of pBQ with that of benzene. Section \ref{sec:conclusion} contains a brief summary.

\section{\label{sec:theory}Theory and numerical implementation}

The positron bound state energy and wavefunction are calculated by solving the Dyson equation,
\begin{equation}\label{dyson_equation}
   \left( \hat{H}_0 + \hat{\Sigma}_{\varepsilon}\right) \psi_{\varepsilon}(\mathbf{r}) = \varepsilon \psi_{\varepsilon}(\mathbf{r}),
\end{equation}
where $\psi_{\varepsilon}(\mathbf{r})$ is the quasiparticle positron wavefunction with energy $\varepsilon$, $\hat{H}_0$ is the Hamiltonian operator of the positron in the static (Hartree-Fock) field of the molecule and $\hat{\Sigma}_{\varepsilon}$ is the positron self-energy operator that accounts for important electron-positron correlation effects. The self-energy operator is nonlocal and dependent on the positron energy $\varepsilon$. In order to obtain a self-consistent solution to the Dyson equation, we solve the equation for a grid of $\varepsilon$ values which spans the positron binding energy and interpolate the data to find where we would have $\varepsilon = \varepsilon_{\rm b}$. A detailed description of the method used to solve this equation is found in Ref. \cite{Hofierka2022}. We employ the fixed-nuclei approximation in these calculations and optimise the molecular geometry at the Hartree-Fock level with aug-cc-pVTZ basis sets \cite{Dunning1992} using the NWChem software \cite{NWChem}.

\begin{figure}[t!]
    \centering
    \includegraphics[width=\linewidth]{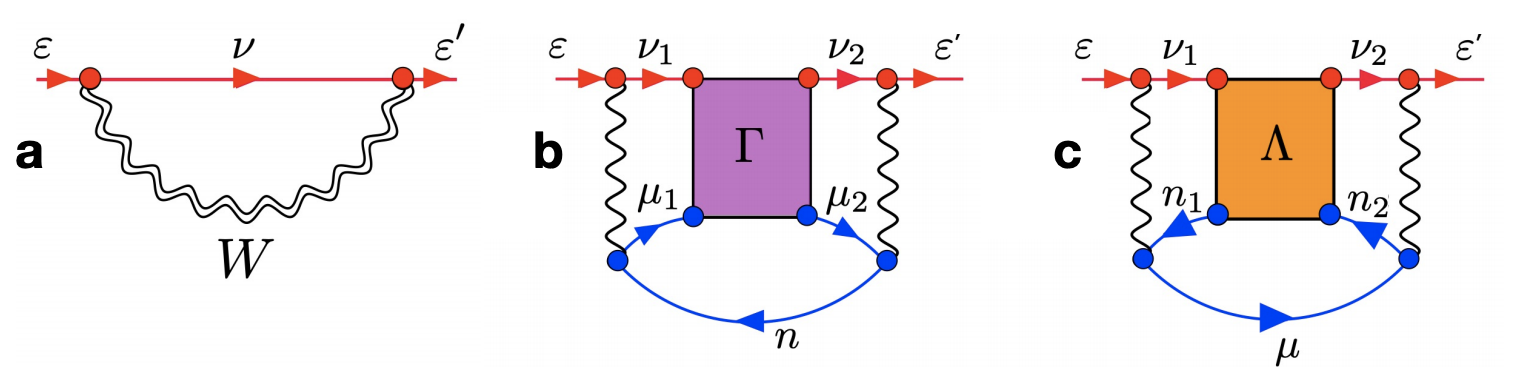}
    \caption{Many-body diagrams for three contributions to the positron self-energy: (a) the $GW$ self-energy which accounts for polarization, screening of the positron-electron Coulomb interaction and electron-hole attraction, (b) the virtual positronium formation diagram, and (c) the positron-hole repulsion diagram.}
    \label{fig:diagrams}
\end{figure}

Our approach accounts for three contributions to the self-energy, represented by the three diagrams in Figure \ref{fig:diagrams}. The first is the $GW$ self-energy $\Sigma^{GW}$, which accounts for polarization of the electron cloud by the positron, screening of the positron-electron Coulomb interaction and electron-hole attraction. Secondly, we calculate the contribution from virtual positronium formation $\Sigma^{\Gamma}$, the process in which a molecular electron temporarily tunnels to the positron to form a positronium-like state. This is depicted diagrammatically by the $\Gamma$ block in Figure \ref{fig:diagrams}(b), which represents an infinite `ladder' of repeated interactions between a positron and an electron. The third, and final contribution to the self-energy $\Sigma^{\Lambda}$ accounts for positron-hole repulsion and also involves an infinite ladder diagram. These three contributions are added to obtain the total self-energy, $\Sigma = \Sigma^{GW} + \Sigma^{\Gamma} + \Sigma^{\Lambda}$.

\begin{figure}[h]
    \centering
    \includegraphics[trim={5cm 6cm 5cm 7cm},clip,width=0.8\linewidth]{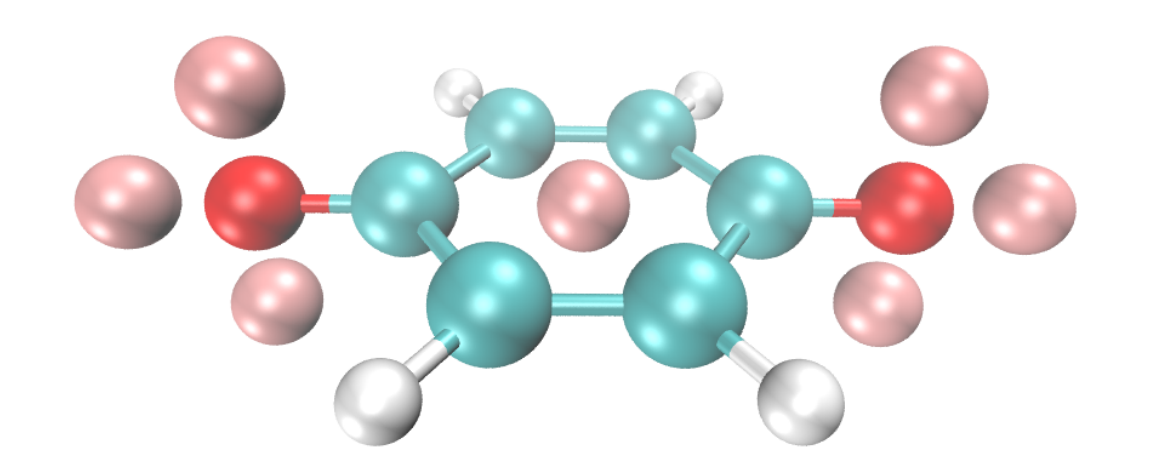}
    \caption{Schematic diagram of the parabenzoquinone molecule showing the locations of additional basis centers. Green (white) ((red)) spheres are carbon (hydrogen) ((oxygen)) atoms, and additional basis centers are marked as pink spheres.}
    \label{fig:ghosts}
\end{figure}

The positron and electron wavefunctions are expanded in Gaussian basis sets, with basis functions placed at each atom in the molecule, and at some additional locations to enhance the basis in regions away from the atomic centers. 
The positron binding energy is converged to within 10\% with respect to changes in the basis and for the final calculation, the basis is as follows. Augmented correlation-consistent polarized valence triple (T) or quadruple (Q) zeta Dunning basis sets \cite{Dunning1992} are placed at each atom for the positron and the electrons: aug-cc-pVTZ basis sets on the C and H atoms and aug-cc-pVQZ basis sets on the O atoms. 
Basis functions are also placed at seven additional locations near the molecule, as shown in Figure \ref{fig:ghosts}. Their locations were chosen by first performing an initial calculation using aug-cc-pVTZ electron and positron bases on every atom (and no additional basis centers) which showed that the positron bound state wavefunction is mostly concentrated in two lobes near the two oxygen atoms. The basis was then enhanced in these regions by incorporating basis sets at three additional centers located 1 $\mathrm{\AA}$ from each oxygen atom, each of which hosts aug-cc-pVTZ hydrogen basis functions for the electron and positron. A set of basis functions is also placed at the center of the molecule's ring which hosts an aug-cc-pVTZ electron basis and a diffuse even-tempered positron basis of the form $10s9p8d7\!f3g$, with smallest exponent $10^{-3}$ and a ratio of 2.2 between consecutive exponents. Including this diffuse positron basis improves the description of long-range correlation effects.

We calculate the electron-positron annihilation contact density $\delta$ using the positron bound-state wavefunction $\psi_{\varepsilon}(\mathbf{r})$ as follows:
\begin{equation}\label{eqn:contact_density}
    \delta = \sum_{n=1}^{N_{\rm e}} \gamma_n \int \left| \phi_n(\mathbf{r})\right|^2\left| \psi_{\varepsilon}(\mathbf{r})\right|^2 d\mathbf{r},
\end{equation}
where $\phi_n(\mathbf{r})$ are the $N_{\rm{e}}$ occupied electronic molecular orbitals ($N_{\rm e}=28$ for pBQ) and $\gamma_n$ are enhancement factors \cite{Green2015} which depend on the $GW$ ionization energy $\varepsilon_n$ of each molecular orbital: $\gamma_n = 1 + \sqrt{1.31/\left|\varepsilon_n\right|} + (0.834 / \left|\varepsilon_n\right|)^{2.15}$. These enhancement factors account for short-range electron-positron attraction \cite{Green2015}.

Contributions to the self-energy are calculated via an expansion in the Hartree-Fock states which is computationally expensive, especially when calculating the virtual positronium formation diagram. In practice, we minimise the computational requirements by neglecting a portion of the highest-energy electron and positron states whilst doing these calculations. In the final calculation for pBQ presented here, we include 588 positron states and 486 electron states (75\% of the total 784 positron and 648 electron states), with energies up to 179 eV and 138 eV, respectively. We note that although the binding energy is quite well-converged with respect to the number of states used in the expansion, complete convergence has not been achieved here due to the computational resources required to perform these calculations, but it is expected that the present binding energy results are within a few meV of a converged value.

All of the calculations in this paper were performed by running the highly parallelized {\tt EXCITON+} code, which was adapted from the {\tt EXCITON} electronic structure code \cite{Patterson2019,Patterson2020} to include positrons, on the UK Tier-2 HPC cluster Kelvin2. The three MBT calculations in Table \ref{tab:results} were performed using 768 processors across 12 nodes, and took between 26 and 38 hours to complete.

\section{\label{sec:results}Results and discussion}
\subsection{Positron binding energy and wavefunction}
\begin{table*}[ht]
\centering
	\caption{\label{tab:results}
        Calculated positron binding energy $\varepsilon_{\rm b}$ and molecular properties for parabenzoquinone and benzene \cite{ArthurBaidoo2024}: dipole moment $\mu$ (at the HF level), dipole polarizability $\alpha$ (at the $GW$@BSE level), ionization energy $I$ (HF (first) and $GW$@RPA (second)) and number of double bonds, $N_{\pi}$. Experimental values \cite{NIST,CRC86} of molecular properties are quoted after calculated values where available. Binding energies are quoted from using three variations of the full MBT calculation, and from using a model to estimate the virtual positronium formation self-energy. Annihilation contact densities $\delta$ are calculated at the $GW{\rm @BSE}+\tilde\Gamma+\tilde\Lambda$ level of MBT with enhancement factors \cite{Green2015}.}   
\begin{tabular}{l@{\hskip12pt}c@{\hskip12pt}c@{\hskip12pt}c@{\hskip12pt}c@{\hskip12pt}l@{\hskip4pt}c@{\hskip12pt}c@{\hskip12pt}c@{\hskip12pt}c@{\hskip12pt}c@{\hskip12pt}}
    \toprule 
    &\multicolumn{4}{c}{\textbf{Molecular properties}}   
    & &\multicolumn{3}{c}{\textbf{$\varepsilon_{\rm b}$ / meV}} \\			
    \cline{2-5}\cline{7-9}
    \\[-1.5ex]
    \textbf{Formula} 
    & $\mu$\,(D) 
    & $\alpha$\,(a.u.) 
    & $I$\,(eV) 
    & $N_{\pi}$ 
    &
    & MBT $^{[1]}$
    & MBT model $^{[2]}$
    & Ref. \cite{Moreira2024}
    &$\delta$ (a.u.) 
    \\
    \hline\\[-1ex]
 \ce{C6H4O2}		& 0.0, 0.0 & 71.9, -- & 11.29, 11.55, 10.11 & 4 &
& 54, 44, \textbf{60} & 46--82 & 0.0925 & 8.00$\times10^{-3}$ \\
 \ce{C6H6}		& 0.0, 0.0 & 66.1, 67.5 & 9.21, 9.57, 9.24 & 3 &
& 158, 122, \textbf{148} & 96--160 & -- & 1.61$\times10^{-2}$ \\[2ex] 
\bottomrule
\end{tabular}
\begin{flushleft}     
[1] Many-body calculations at three levels of $\Sigma^{GW+\Gamma+\Lambda}$. The first (second) number uses bare (dressed) Coulomb interactions within the ladders and HF energies. The third, highlighted in bold, is our most sophisticated calculation which uses dressed Coulomb interactions and $GW$@RPA energies in the ladders.

[2] Using Equation \ref{eq:mbtmodel} to approximate the self-energy. Values of $g=1.4$ ($g=1.5$) are used to give the lower (upper) bound of the range.
\end{flushleft} 
\end{table*}

This section presents results from our many-body theory positron binding calculations for pBQ. Calculated positron binding energies are quoted in Table \ref{tab:results} alongside calculated values of the dipole moment, polarizability and ionization energy. Results for benzene from Ref. \cite{ArthurBaidoo2024} are also shown for comparison.

Table \ref{tab:results} quotes three results for the positron-pBQ binding energy calculated using the full $GW+\Gamma+\Lambda$ self-energy. These three results are obtained by using different methods to evaluate the ladder diagrams from Figure \ref{fig:diagrams}(b) and (c), as described in the note below the table. Our most sophisticated ($GW+\tilde\Gamma+\tilde\Lambda$) calculation gives a positron-pBQ binding energy of $\varepsilon_{\rm b}$ = 60 meV. An error bar is placed on this binding energy using the difference between the largest and smallest of the three results, so that our final calculated binding energy is given as 60 $\pm$ 16 meV. 
Binding calculations were also carried out at the Hartree-Fock level of theory, where none of the self-energy contributions in Figure \ref{fig:diagrams} are included, and at the $GW$@BSE level, where only the diagram in Figure \ref{fig:diagrams}(a) is included,  but neither of these approaches predicted any positron binding to the pBQ molecule, highlighting the importance of accounting for all of the correlation effects in Figure \ref{fig:diagrams}. Our calculated positron-pBQ binding energy is much larger than the value of $\varepsilon_{\rm b}=$ 0.0925 meV obtained in Ref. \cite{Moreira2024}, but this is expected since our approach includes additional correlation effects which enhance binding. 

In addition to the full many-body theory results, an estimate for the positron-pBQ binding energy was obtained using a model \cite{Hofierka2022} which approximates the virtual positronium formation self-energy by scaling the bare polarization contribution to the self-energy, $\Sigma^{(2)}$, by a parameter $g$, so that the self-energy is calculated as
\begin{equation}\label{eq:mbtmodel}
    \Sigma \approx g \Sigma^{(2)} + \Sigma^{\Lambda}.
\end{equation}
By avoiding the computationally-intensive calculation of the virtual positronium formation self-energy, this model is an efficient method of estimating the positron binding energy. Typically, setting $g$ = 1.4 gives a lower estimate of the binding energy and setting $g$ = 1.5 gives an upper estimate (see e.g., \cite{Hofierka2022,Cassidy2023} for examples of binding energies calculated with the model compared with the fully \emph{ab initio} description of the self energy). For pBQ, the model gives 46 meV $\leq \varepsilon_{\rm b} \leq$  82 meV which is in good agreement with the full MBT result.

\begin{figure}
    \centering
    \includegraphics[trim={1.5cm 9cm 4.5cm 8cm},clip,width=\linewidth]{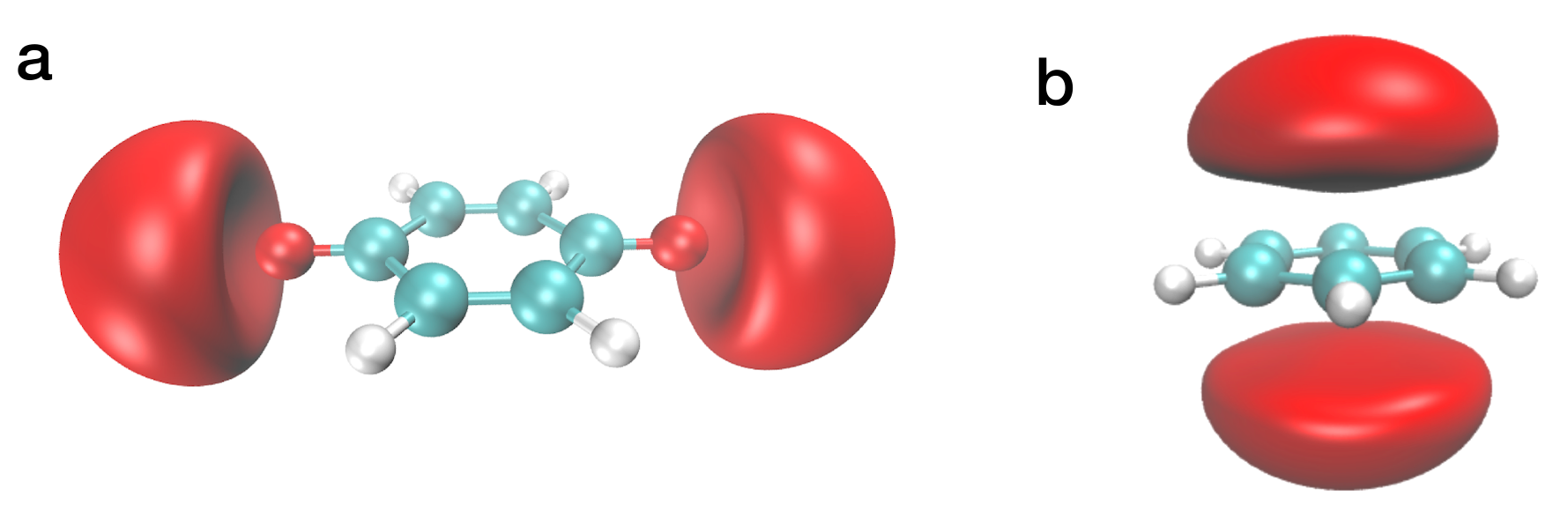}
    \caption{The lowest positron bound states (Dyson orbitals) in (a) parabenzoquinone and (b) benzene \cite{ArthurBaidoo2024}; green (white) ((red)) spheres are carbon (hydrogen) ((oxygen)) atoms, solid red isosurfaces are the positron Dyson wavefunctions at 80\% of their maximum values. Also see Fig.~\ref{fig:electron_density} for contour plots. }
    \label{fig:positron_wfn}
\end{figure}

\begin{figure*}
    \centering
    \includegraphics[trim={1cm 5cm 1cm 1cm}, clip, width=0.98\linewidth]{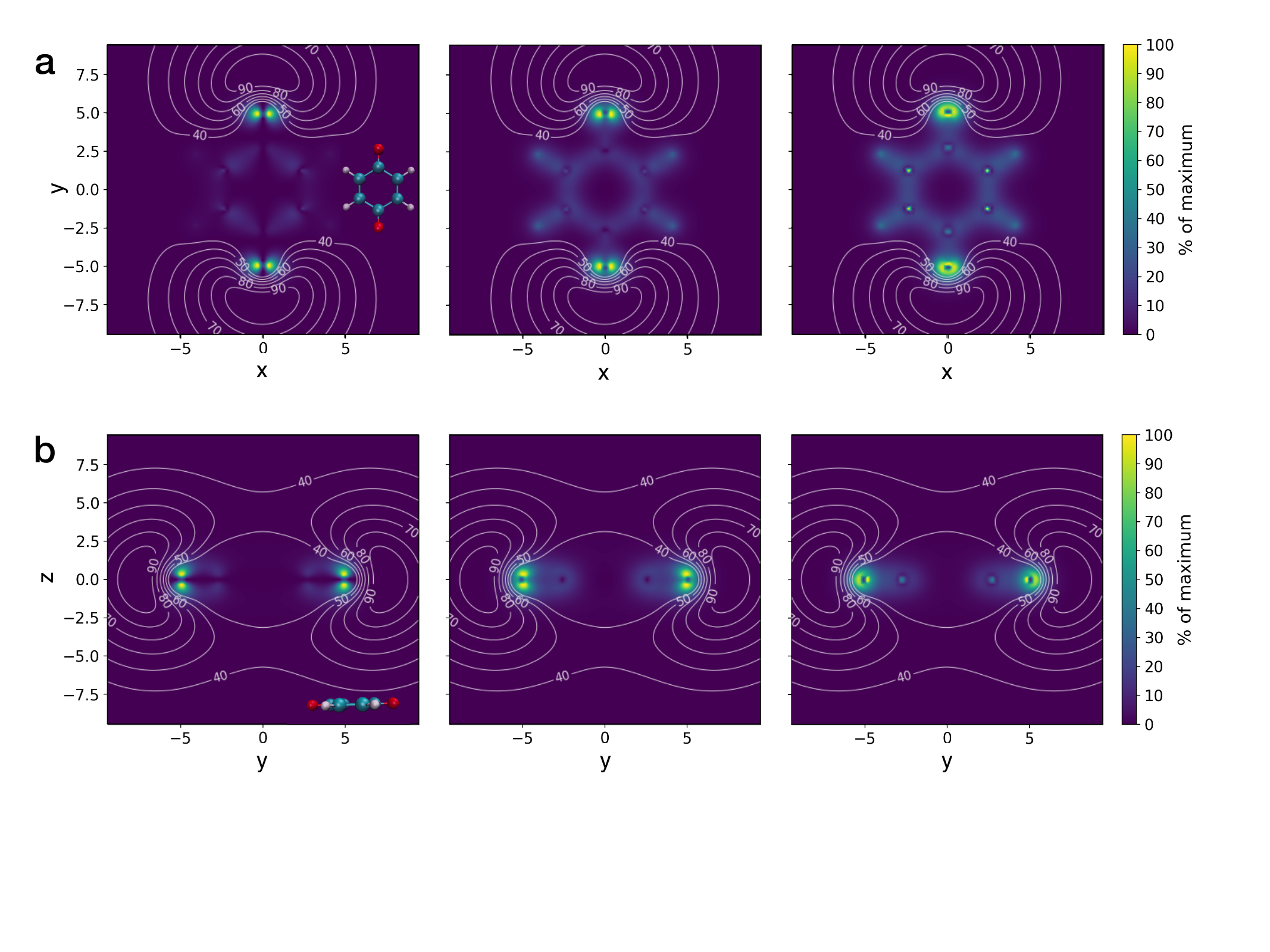}
    \caption{Plots of the electron and positron density in the parabenzoquinone molecule (a) in the $xy$ plane ($z=0$) and (b) in the $yz$ plane ($x=0$). The molecular geometry is such that the centre of the parabenzoquinone ring is at the origin of the coordinate space and inset images show the orientation of the molecule in the $xy$ and $yz$ planes. The electron density is shown by the colour map plot and the left, middle and right panels show the combined electron density from the top 5, 10 and 15 molecular orbitals, respectively. Contour lines represent the positron density at 40\%-90\% of the maximum value in steps of 10\%. In the $xy$-plane plot for the top 5 MOs, it is noted that only the (H-2)OMO and (H-3)OMO make contributions to the electron density, since the other three MOs have a node in the electron density in this plane.}
    \label{fig:electron_density}
\end{figure*}

The positron bound-state wavefunction, or Dyson orbital, in pBQ is shown in Figure \ref{fig:positron_wfn}(a) as a solid red surface, which is an isosurface of the wavefunction at 80\% of its maximum value. 
The positron density is strongly localized to the two regions next to the oxygen atoms at opposite sides of the molecule, indicating a strong attraction of the positron to these two atoms. 
A molecule's dipole moment typically plays a significant role in determining the location of the positron Dyson orbital, with the positron density usually concentrated at the negative side of the molecule, but since pBQ has a zero dipole moment, this effect is absent and the positron density is distributed evenly at both sides of the molecule, in keeping with the molecule's $D_{2\rm h}$ symmetry. 
Although the pBQ molecule itself has zero dipole moment, the C=O (carbonyl) bonds at opposite sides of the ring are polarized such that there is a higher negative charge density at the O atoms than the C atoms, thus making the O atoms more attractive to the positron. 
Additionally, the O atoms in pBQ each have two lone pairs of electrons which attract the positron towards them. 
Figure \ref{fig:electron_density} shows the high electron density around the O atoms alongside a contour plot of the positron density, which is strongly localised near these regions of high negative charge.
Double ($\pi$) bonds in a molecule also attract the positron due to their electron density being mostly out of the plane of the molecule, away from the positive atomic nuclei, which the positron cannot penetrate close to. The pBQ molecule has a total of four $\pi$ bonds; two in the ring of the molecule and two carbonyl bonds at opposite sides of the molecule. 
It is evident from Figure \ref{fig:electron_density}(b) that there is a slightly increased positron density in the regions just above and below the plane of the molecule, indicating some attraction to the electrons from $\pi$ bonds in the pBQ ring.

We compare the calculated positron bound state in pBQ to that obtained for benzene (\ce{C6H6}) with the present MBT methods. The calculated positron binding energy for benzene is 148$\pm$26 meV, in agreement with the experimental value of 132$\pm$3 meV from the same study \cite{ArthurBaidoo2024}. Thus, replacing two of the H atoms with O atoms to form pBQ significantly reduces the binding energy. 
The shape of the pBQ Dyson orbital is also very different to the positron bound state wavefunction for the benzene molecule, shown in Figure \ref{fig:positron_wfn}(b), which has its regions of high positron density above and below the molecule's ring.  One key difference between benzene and pBQ is that benzene is aromatic, while pBQ is not: in benzene, there is a region of high electron density above and below the molecular plane due to the aromatic ring which attracts the positron to these regions, whereas pBQ does not have a complete aromatic ring and the positron instead favours binding to the O atoms. 
The smaller positron binding energy in pBQ compared with benzene can be ascribed to the loss of aromaticity; the electron density near the O atoms is closer to the repulsive oxygen nuclei and thus, more difficult for the positron to probe compared with the electron density above and below the ring of the benzene molecule.

\subsection{Molecular orbital contributions to the positron-molecule correlation potential}

\begin{figure*}
    \centering
    \includegraphics[trim={0cm 4.5cm 1.8cm 4.5cm},clip,width=\linewidth]{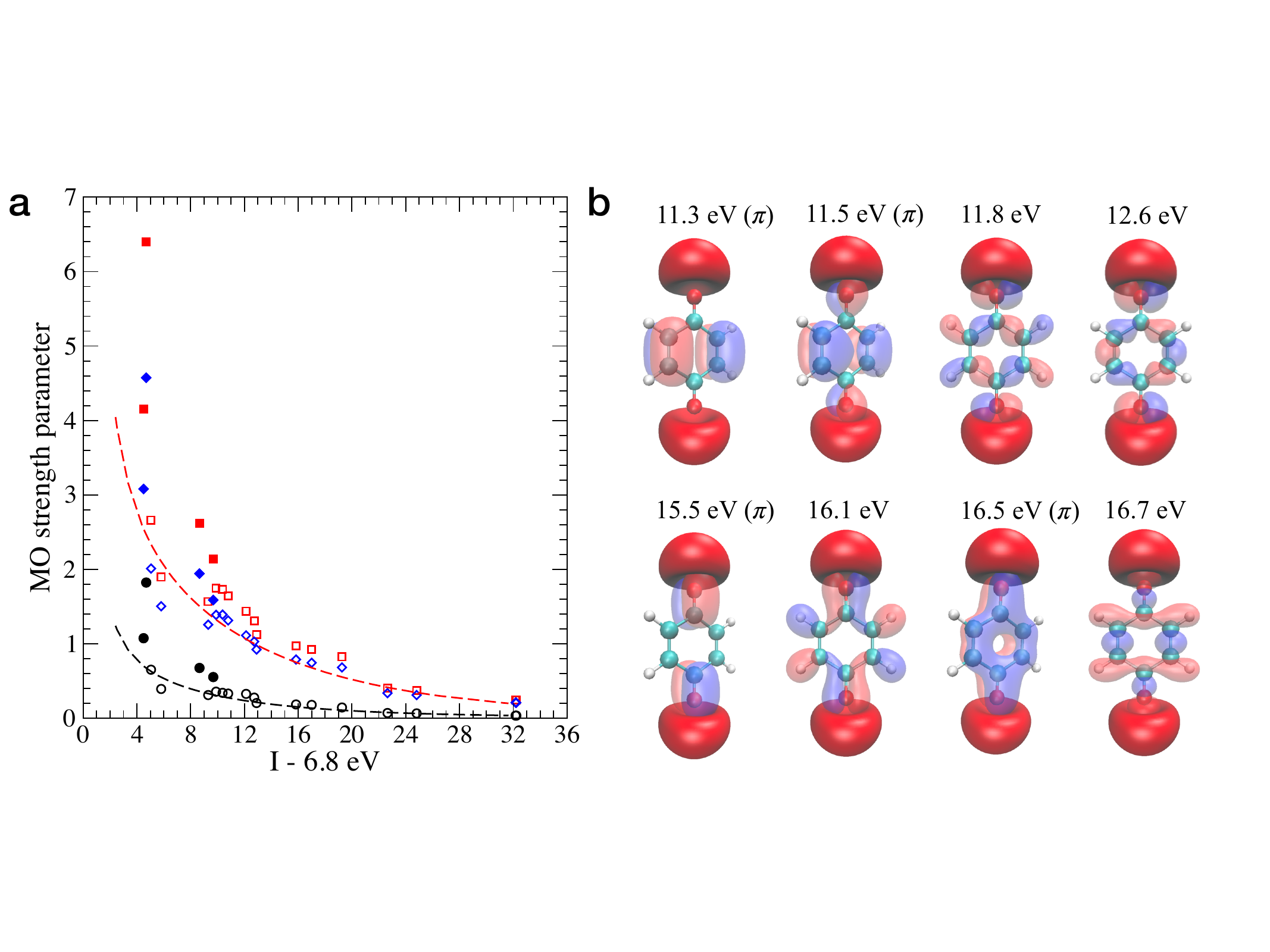}
    \caption{(a) Strength parameters for the electron molecular orbitals of parabenzoquinone plotted against the Hartree-Fock ionization energy with the positronium formation threshold, 6.8eV, subtracted. Blue diamonds: $S^{(2)}$, black circles: $S^{(\Gamma)}$, red squares: $S^{(2+\Gamma)}$. Strength parameters for $\pi$ type orbitals are distinguished by solid markers. Dashed lines show the fits from Ref. \cite{Hofierka2022}, Figure 3. (b) The eight highest occupied electron molecular orbitals of parabenzoquinone, plotted as isosurfaces at values of $\pm$0.04 (transparent blue lobes are negative valued and transparent red lobes are positive valued). Ionization energies are given above each image and $\pi$ orbitals are labelled. The positron Dyson orbital is plotted as an opaque red isosurface at 80\% of its maximum value.
    }
    \label{fig:electron_orbitals_str_param}
\end{figure*}

Contributions to the self-energy from individual molecular orbitals can be quantified using strength parameters $S$ which are calculated as follows \cite{Dzuba1994}:
\begin{equation}
    S = -\sum_{\nu>0}\varepsilon_{\nu}^{-1} \langle \nu| \Sigma |\nu \rangle,
\end{equation}
where $\Sigma$ is the self-energy and the sum runs over excited Hartree-Fock states with $\varepsilon_{\nu}>0$. Strength parameters are calculated for each molecular orbital, with a larger strength parameter indicating a stronger contribution to the self-energy and thus, to binding.

Figure \ref{fig:electron_orbitals_str_param}(a) shows calculated strength parameters $S$ for pBQ plotted against the ionization energy for each MO with the positronium formation threshold subtracted. The tightly-bound core orbitals have been omitted from the plot as they make minimal contributions to binding, having almost-zero strength parameters. Typically, the contributions of MOs to the positron-molecule correlation potential are strongly dependent on the MO ionization energy: more tightly-bound electrons have less influence on the positron since they are close to the positive atomic nuclei which repel the positron. 

It is understood that the positron binding energy to molecules is correlated positively with the number of $\pi$ bonds in the molecule \cite{Danielson09,Suguira2020,Hofierka2022, ArthurBaidoo2024}, and exceptions to the ordering of strength parameters with ionization energy are found in the $\pi$ orbitals.
The eight highest occupied electron molecular orbitals of pBQ are depicted in Figure \ref{fig:electron_orbitals_str_param}(b) alongside the positron Dyson orbital, and the four $\pi$-type orbitals for pBQ (corresponding to the four double bonds) are easily identified by their shape: they have a node in the plane of the molecule, and reflective symmetry in that plane. From Figure \ref{fig:electron_orbitals_str_param}(b), we can see that the $\pi$ orbitals are those with energies 11.3 eV, 11.5 eV, 15.5 eV and 16.5 eV. Figure \ref{fig:electron_orbitals_str_param}(a) shows that the four $\pi$ orbitals (denoted by solid symbols) have four of the five largest strength parameters at all of the three levels of theory shown, despite other orbitals having lower ionization energies.
Looking at the HOMO (binding energy 11.3 eV) and the (H–1)OMO (binding energy 11.5 eV), we see that the strength parameters for the (H–1)OMO are much larger than those for the HOMO, even though both molecular orbitals have similar energies and $\pi$-type symmetry. This difference can be understood using the diagrams in Figure \ref{fig:electron_orbitals_str_param}(b); the electron density in the (H–1)OMO overlaps with the positron wavefunction more than that of the HOMO due to its lobes of high electron density surrounding the O atoms and thus, the shape of the (H–1)OMO favours interaction between the positron and electrons in this molecular orbital.

\section{Summary}\label{sec:conclusion}
Our many-body theory calculations predict that a positron can bind to the parabenzoquinone molecule with a binding energy of $\varepsilon_{\rm b} = 60 \pm 16$ meV. This result is much lower than the calculated binding energy for benzene, $\varepsilon_{\rm b} = 148\pm26 $ meV, which is attributed to the fact that compared to benzene, pBQ is not aromatic: rather than being free to probe the electron density in the aromatic rings which are out of the plane away from the positive nuclei, in pBQ the positron is localized in regions of significant electron density near the oxygen nuclei. 
The strength of contribution to the positron-molecule correlation potential from molecular orbitals of different symmetry was studied:  the four $\pi$ orbitals contribute strongly to binding, particularly the (H-1)OMO which also has high electron density next to the oxygen atoms. The positron binding energy calculated in this work is much larger than that obtained from the scattering calculations in Ref. \cite{Moreira2024}. It would thus be instructive to study the scattering properties of pBQ using the positron-molecule self-energy employed in this work. Efforts to develop a computational \emph{ab initio} many-body theory approach to positron-molecule scattering that may allow this are underway. 

\section{Acknowledgements}\label{sec:ack}
We thank James Danielson (UCSD) for useful discussions.


\begin{thebibliography}{25}%
\makeatletter
\providecommand \@ifxundefined [1]{%
 \@ifx{#1\undefined}
}%
\providecommand \@ifnum [1]{%
 \ifnum #1\expandafter \@firstoftwo
 \else \expandafter \@secondoftwo
 \fi
}%
\providecommand \@ifx [1]{%
 \ifx #1\expandafter \@firstoftwo
 \else \expandafter \@secondoftwo
 \fi
}%
\providecommand \natexlab [1]{#1}%
\providecommand \enquote  [1]{``#1''}%
\providecommand \bibnamefont  [1]{#1}%
\providecommand \bibfnamefont [1]{#1}%
\providecommand \citenamefont [1]{#1}%
\providecommand \href@noop [0]{\@secondoftwo}%
\providecommand \href [0]{\begingroup \@sanitize@url \@href}%
\providecommand \@href[1]{\@@startlink{#1}\@@href}%
\providecommand \@@href[1]{\endgroup#1\@@endlink}%
\providecommand \@sanitize@url [0]{\catcode `\\12\catcode `\$12\catcode
  `\&12\catcode `\#12\catcode `\^12\catcode `\_12\catcode `\%12\relax}%
\providecommand \@@startlink[1]{}%
\providecommand \@@endlink[0]{}%
\providecommand \url  [0]{\begingroup\@sanitize@url \@url }%
\providecommand \@url [1]{\endgroup\@href {#1}{\urlprefix }}%
\providecommand \urlprefix  [0]{URL }%
\providecommand \Eprint [0]{\href }%
\providecommand \doibase [0]{https://doi.org/}%
\providecommand \selectlanguage [0]{\@gobble}%
\providecommand \bibinfo  [0]{\@secondoftwo}%
\providecommand \bibfield  [0]{\@secondoftwo}%
\providecommand \translation [1]{[#1]}%
\providecommand \BibitemOpen [0]{}%
\providecommand \bibitemStop [0]{}%
\providecommand \bibitemNoStop [0]{.\EOS\space}%
\providecommand \EOS [0]{\spacefactor3000\relax}%
\providecommand \BibitemShut  [1]{\csname bibitem#1\endcsname}%
\let\auto@bib@innerbib\@empty
\bibitem [{\citenamefont {Gilbert}\ \emph {et~al.}(2002)\citenamefont
  {Gilbert}, \citenamefont {Barnes}, \citenamefont {Sullivan},\ and\
  \citenamefont {Surko}}]{Gilbert02}%
  \BibitemOpen
  \bibfield  {author} {\bibinfo {author} {\bibfnamefont {S.~J.}\ \bibnamefont
  {Gilbert}}, \bibinfo {author} {\bibfnamefont {L.~D.}\ \bibnamefont {Barnes}},
  \bibinfo {author} {\bibfnamefont {J.~P.}\ \bibnamefont {Sullivan}},\ and\
  \bibinfo {author} {\bibfnamefont {C.~M.}\ \bibnamefont {Surko}},\ }\bibfield
  {title} {\bibinfo {title} {Vibrational-resonance enhancement of positron
  annihilation in molecules},\ }\href
  {https://doi.org/10.1103/PhysRevLett.88.043201} {\bibfield  {journal}
  {\bibinfo  {journal} {Phys. Rev. Lett.}\ }\textbf {\bibinfo {volume} {88}},\
  \bibinfo {pages} {043201} (\bibinfo {year} {2002})}\BibitemShut {NoStop}%
\bibitem [{\citenamefont {Danielson}\ \emph {et~al.}(2009)\citenamefont
  {Danielson}, \citenamefont {Young},\ and\ \citenamefont
  {Surko}}]{Danielson09}%
  \BibitemOpen
  \bibfield  {author} {\bibinfo {author} {\bibfnamefont {J.~R.}\ \bibnamefont
  {Danielson}}, \bibinfo {author} {\bibfnamefont {J.~A.}\ \bibnamefont
  {Young}},\ and\ \bibinfo {author} {\bibfnamefont {C.~M.}\ \bibnamefont
  {Surko}},\ }\bibfield  {title} {\bibinfo {title} {Dependence of
  positron-molecule binding energies on molecular properties},\ }\href
  {https://doi.org/10.1088/0953-4075/42/23/235203} {\bibfield  {journal}
  {\bibinfo  {journal} {J. Phys. B}\ }\textbf {\bibinfo {volume} {42}},\
  \bibinfo {pages} {235203} (\bibinfo {year} {2009})}\BibitemShut {NoStop}%
\bibitem [{\citenamefont {Danielson}\ \emph {et~al.}(2010)\citenamefont
  {Danielson}, \citenamefont {Gosselin},\ and\ \citenamefont
  {Surko}}]{Danielson10}%
  \BibitemOpen
  \bibfield  {author} {\bibinfo {author} {\bibfnamefont {J.~R.}\ \bibnamefont
  {Danielson}}, \bibinfo {author} {\bibfnamefont {J.~J.}\ \bibnamefont
  {Gosselin}},\ and\ \bibinfo {author} {\bibfnamefont {C.~M.}\ \bibnamefont
  {Surko}},\ }\bibfield  {title} {\bibinfo {title} {Dipole enhancement of
  positron binding to molecules},\ }\href
  {https://doi.org/10.1103/PhysRevLett.104.233201} {\bibfield  {journal}
  {\bibinfo  {journal} {Phys. Rev. Lett.}\ }\textbf {\bibinfo {volume} {104}},\
  \bibinfo {pages} {233201} (\bibinfo {year} {2010})}\BibitemShut {NoStop}%
\bibitem [{\citenamefont {Danielson}\ \emph
  {et~al.}(2012{\natexlab{a}})\citenamefont {Danielson}, \citenamefont {Jones},
  \citenamefont {Gosselin}, \citenamefont {Natisin},\ and\ \citenamefont
  {Surko}}]{Danielson12}%
  \BibitemOpen
  \bibfield  {author} {\bibinfo {author} {\bibfnamefont {J.~R.}\ \bibnamefont
  {Danielson}}, \bibinfo {author} {\bibfnamefont {A.~C.~L.}\ \bibnamefont
  {Jones}}, \bibinfo {author} {\bibfnamefont {J.~J.}\ \bibnamefont {Gosselin}},
  \bibinfo {author} {\bibfnamefont {M.~R.}\ \bibnamefont {Natisin}},\ and\
  \bibinfo {author} {\bibfnamefont {C.~M.}\ \bibnamefont {Surko}},\ }\bibfield
  {title} {\bibinfo {title} {Interplay between permanent dipole moments and
  polarizability in positron-molecule binding},\ }\href
  {https://doi.org/10.1103/PhysRevA.85.022709} {\bibfield  {journal} {\bibinfo
  {journal} {Phys. Rev. A}\ }\textbf {\bibinfo {volume} {85}},\ \bibinfo
  {pages} {022709} (\bibinfo {year} {2012}{\natexlab{a}})}\BibitemShut
  {NoStop}%
\bibitem [{\citenamefont {Danielson}\ \emph
  {et~al.}(2012{\natexlab{b}})\citenamefont {Danielson}, \citenamefont {Jones},
  \citenamefont {Natisin},\ and\ \citenamefont {Surko}}]{Danielson12a}%
  \BibitemOpen
  \bibfield  {author} {\bibinfo {author} {\bibfnamefont {J.~R.}\ \bibnamefont
  {Danielson}}, \bibinfo {author} {\bibfnamefont {A.~C.~L.}\ \bibnamefont
  {Jones}}, \bibinfo {author} {\bibfnamefont {M.~R.}\ \bibnamefont {Natisin}},\
  and\ \bibinfo {author} {\bibfnamefont {C.~M.}\ \bibnamefont {Surko}},\
  }\bibfield  {title} {\bibinfo {title} {Comparisons of {P}ositron and
  {E}lectron {B}inding to {M}olecules},\ }\href
  {https://doi.org/10.1103/PhysRevLett.109.113201} {\bibfield  {journal}
  {\bibinfo  {journal} {Phys. Rev. Lett.}\ }\textbf {\bibinfo {volume} {109}},\
  \bibinfo {pages} {113201} (\bibinfo {year} {2012}{\natexlab{b}})}\BibitemShut
  {NoStop}%
\bibitem [{\citenamefont {Arthur-Baidoo}\ \emph {et~al.}(2024)\citenamefont
  {Arthur-Baidoo}, \citenamefont {Danielson}, \citenamefont {Surko},
  \citenamefont {Cassidy}, \citenamefont {Gregg}, \citenamefont {Hofierka},
  \citenamefont {Cunningham}, \citenamefont {Patterson},\ and\ \citenamefont
  {Green}}]{ArthurBaidoo2024}%
  \BibitemOpen
  \bibfield  {author} {\bibinfo {author} {\bibfnamefont {E.}~\bibnamefont
  {Arthur-Baidoo}}, \bibinfo {author} {\bibfnamefont {J.~R.}\ \bibnamefont
  {Danielson}}, \bibinfo {author} {\bibfnamefont {C.~M.}\ \bibnamefont
  {Surko}}, \bibinfo {author} {\bibfnamefont {J.~P.}\ \bibnamefont {Cassidy}},
  \bibinfo {author} {\bibfnamefont {S.~K.}\ \bibnamefont {Gregg}}, \bibinfo
  {author} {\bibfnamefont {J.}~\bibnamefont {Hofierka}}, \bibinfo {author}
  {\bibfnamefont {B.}~\bibnamefont {Cunningham}}, \bibinfo {author}
  {\bibfnamefont {C.~H.}\ \bibnamefont {Patterson}},\ and\ \bibinfo {author}
  {\bibfnamefont {D.~G.}\ \bibnamefont {Green}},\ }\bibfield  {title} {\bibinfo
  {title} {Positron annihilation and binding in aromatic and other ring
  molecules},\ }\href {https://doi.org/10.1103/PhysRevA.109.062801} {\bibfield
  {journal} {\bibinfo  {journal} {Phys. Rev. A}\ }\textbf {\bibinfo {volume}
  {109}},\ \bibinfo {pages} {062801} (\bibinfo {year} {2024})}\BibitemShut
  {NoStop}%
\bibitem [{\citenamefont {Tachikawa}\ \emph {et~al.}(2011)\citenamefont
  {Tachikawa}, \citenamefont {Kita},\ and\ \citenamefont
  {Buenker}}]{Tachikawa2011}%
  \BibitemOpen
  \bibfield  {author} {\bibinfo {author} {\bibfnamefont {M.}~\bibnamefont
  {Tachikawa}}, \bibinfo {author} {\bibfnamefont {Y.}~\bibnamefont {Kita}},\
  and\ \bibinfo {author} {\bibfnamefont {R.~J.}\ \bibnamefont {Buenker}},\
  }\bibfield  {title} {\bibinfo {title} {Bound states of the positron with
  nitrile species with a configuration interaction multi-component molecular
  orbital approach},\ }\href {https://doi.org/10.1039/C0CP01650K} {\bibfield
  {journal} {\bibinfo  {journal} {Phys. Chem. Chem. Phys.}\ }\textbf {\bibinfo
  {volume} {13}},\ \bibinfo {pages} {2701} (\bibinfo {year}
  {2011})}\BibitemShut {NoStop}%
\bibitem [{\citenamefont {Koyanagi}\ \emph {et~al.}(2012)\citenamefont
  {Koyanagi}, \citenamefont {Kita},\ and\ \citenamefont
  {Tachikawa}}]{Koyanagi2012}%
  \BibitemOpen
  \bibfield  {author} {\bibinfo {author} {\bibfnamefont {K.}~\bibnamefont
  {Koyanagi}}, \bibinfo {author} {\bibfnamefont {Y.}~\bibnamefont {Kita}},\
  and\ \bibinfo {author} {\bibfnamefont {M.}~\bibnamefont {Tachikawa}},\
  }\bibfield  {title} {\bibinfo {title} {Systematic theoretical investigation
  of a positron binding to amino acid molecules using the ab initio
  multi-component molecular orbital approach.},\ }\href@noop {} {\bibfield
  {journal} {\bibinfo  {journal} {Eur. Phys. J. D}\ }\textbf {\bibinfo {volume}
  {66}},\ \bibinfo {pages} {121} (\bibinfo {year} {2012})}\BibitemShut
  {NoStop}%
\bibitem [{\citenamefont {Charry}\ \emph {et~al.}(2014)\citenamefont {Charry},
  \citenamefont {Romero}, \citenamefont {Varella},\ and\ \citenamefont
  {Reyes}}]{Charry2014}%
  \BibitemOpen
  \bibfield  {author} {\bibinfo {author} {\bibfnamefont {J.}~\bibnamefont
  {Charry}}, \bibinfo {author} {\bibfnamefont {J.}~\bibnamefont {Romero}},
  \bibinfo {author} {\bibfnamefont {M.~T. d.~N.}\ \bibnamefont {Varella}},\
  and\ \bibinfo {author} {\bibfnamefont {A.}~\bibnamefont {Reyes}},\ }\bibfield
   {title} {\bibinfo {title} {Calculation of positron binding energies of amino
  acids with the any-particle molecular-orbital approach},\ }\href
  {https://doi.org/10.1103/PhysRevA.89.052709} {\bibfield  {journal} {\bibinfo
  {journal} {Phys. Rev. A}\ }\textbf {\bibinfo {volume} {89}},\ \bibinfo
  {pages} {052709} (\bibinfo {year} {2014})}\BibitemShut {NoStop}%
\bibitem [{\citenamefont {Swann}\ and\ \citenamefont
  {Gribakin}(2019)}]{Swann2019}%
  \BibitemOpen
  \bibfield  {author} {\bibinfo {author} {\bibfnamefont {A.~R.}\ \bibnamefont
  {Swann}}\ and\ \bibinfo {author} {\bibfnamefont {G.~F.}\ \bibnamefont
  {Gribakin}},\ }\bibfield  {title} {\bibinfo {title} {Positron binding and
  annihilation in alkane molecules},\ }\href
  {https://doi.org/10.1103/PhysRevLett.123.113402} {\bibfield  {journal}
  {\bibinfo  {journal} {Phys. Rev. Lett.}\ }\textbf {\bibinfo {volume} {123}},\
  \bibinfo {pages} {113402} (\bibinfo {year} {2019})}\BibitemShut {NoStop}%
\bibitem [{\citenamefont {Sugiura}\ \emph {et~al.}(2020)\citenamefont
  {Sugiura}, \citenamefont {Suzuki}, \citenamefont {Otomo}, \citenamefont
  {Miyazaki}, \citenamefont {Takayanagi},\ and\ \citenamefont
  {Tachikawa}}]{Suguira2020}%
  \BibitemOpen
  \bibfield  {author} {\bibinfo {author} {\bibfnamefont {Y.}~\bibnamefont
  {Sugiura}}, \bibinfo {author} {\bibfnamefont {H.}~\bibnamefont {Suzuki}},
  \bibinfo {author} {\bibfnamefont {T.}~\bibnamefont {Otomo}}, \bibinfo
  {author} {\bibfnamefont {T.}~\bibnamefont {Miyazaki}}, \bibinfo {author}
  {\bibfnamefont {T.}~\bibnamefont {Takayanagi}},\ and\ \bibinfo {author}
  {\bibfnamefont {M.}~\bibnamefont {Tachikawa}},\ }\bibfield  {title} {\bibinfo
  {title} {Positron-electron correlation-polarization potential model for
  positron binding in polyatomic molecules},\ }\href
  {https://doi.org/10.1002/jcc.26200} {\bibfield  {journal} {\bibinfo
  {journal} {Journal of computational chemistry}\ }\textbf {\bibinfo {volume}
  {41}},\ \bibinfo {pages} {1576} (\bibinfo {year} {2020})}\BibitemShut
  {NoStop}%
\bibitem [{\citenamefont {Hofierka}\ \emph {et~al.}(2022)\citenamefont
  {Hofierka}, \citenamefont {Cunningham}, \citenamefont {Rawlins},
  \citenamefont {Patterson},\ and\ \citenamefont {Green}}]{Hofierka2022}%
  \BibitemOpen
  \bibfield  {author} {\bibinfo {author} {\bibfnamefont {J.}~\bibnamefont
  {Hofierka}}, \bibinfo {author} {\bibfnamefont {B.}~\bibnamefont
  {Cunningham}}, \bibinfo {author} {\bibfnamefont {C.~M.}\ \bibnamefont
  {Rawlins}}, \bibinfo {author} {\bibfnamefont {C.~H.}\ \bibnamefont
  {Patterson}},\ and\ \bibinfo {author} {\bibfnamefont {D.~G.}\ \bibnamefont
  {Green}},\ }\bibfield  {title} {\bibinfo {title} {Many-body theory of
  positron binding to polyatomic molecules},\ }\href@noop {} {\bibfield
  {journal} {\bibinfo  {journal} {Nature}\ }\textbf {\bibinfo {volume} {606}},\
  \bibinfo {pages} {688} (\bibinfo {year} {2022})}\BibitemShut {NoStop}%
\bibitem [{\citenamefont {Cassidy}\ \emph
  {et~al.}(2024{\natexlab{a}})\citenamefont {Cassidy}, \citenamefont
  {Hofierka}, \citenamefont {Cunningham}, \citenamefont {Rawlins},
  \citenamefont {Patterson},\ and\ \citenamefont {Green}}]{Cassidy2023}%
  \BibitemOpen
  \bibfield  {author} {\bibinfo {author} {\bibfnamefont {J.~P.}\ \bibnamefont
  {Cassidy}}, \bibinfo {author} {\bibfnamefont {J.}~\bibnamefont {Hofierka}},
  \bibinfo {author} {\bibfnamefont {B.}~\bibnamefont {Cunningham}}, \bibinfo
  {author} {\bibfnamefont {C.~M.}\ \bibnamefont {Rawlins}}, \bibinfo {author}
  {\bibfnamefont {C.~H.}\ \bibnamefont {Patterson}},\ and\ \bibinfo {author}
  {\bibfnamefont {D.~G.}\ \bibnamefont {Green}},\ }\bibfield  {title} {\bibinfo
  {title} {Many-body theory calculations of positron binding to halogenated
  hydrocarbons},\ }\href {https://doi.org/10.1103/PhysRevA.109.L040801}
  {\bibfield  {journal} {\bibinfo  {journal} {Phys. Rev. A}\ }\textbf {\bibinfo
  {volume} {109}},\ \bibinfo {pages} {L040801} (\bibinfo {year}
  {2024}{\natexlab{a}})}\BibitemShut {NoStop}%
\bibitem [{\citenamefont {Hofierka}\ \emph {et~al.}(2024)\citenamefont
  {Hofierka}, \citenamefont {Cunningham},\ and\ \citenamefont
  {Green}}]{Hofierka2024}%
  \BibitemOpen
  \bibfield  {author} {\bibinfo {author} {\bibfnamefont {J.}~\bibnamefont
  {Hofierka}}, \bibinfo {author} {\bibfnamefont {B.}~\bibnamefont
  {Cunningham}},\ and\ \bibinfo {author} {\bibfnamefont {D.~G.}\ \bibnamefont
  {Green}},\ }\bibfield  {title} {\bibinfo {title} {Many-body theory
  calculations of positron binding to hydrogen cyanide},\ }\bibfield  {journal}
  {\bibinfo  {journal} {Eur. Phys. J. D}\ }\textbf {\bibinfo {volume} {78}},\
  \href {https://doi.org/10.1140/epjd/s10053-024-00810-0}
  {10.1140/epjd/s10053-024-00810-0} (\bibinfo {year} {2024})\BibitemShut
  {NoStop}%
\bibitem [{\citenamefont {Rawlins}\ \emph {et~al.}(2023)\citenamefont
  {Rawlins}, \citenamefont {Hofierka}, \citenamefont {Cunningham},
  \citenamefont {Patterson},\ and\ \citenamefont {Green}}]{Rawlins2023}%
  \BibitemOpen
  \bibfield  {author} {\bibinfo {author} {\bibfnamefont {C.~M.}\ \bibnamefont
  {Rawlins}}, \bibinfo {author} {\bibfnamefont {J.}~\bibnamefont {Hofierka}},
  \bibinfo {author} {\bibfnamefont {B.}~\bibnamefont {Cunningham}}, \bibinfo
  {author} {\bibfnamefont {C.~H.}\ \bibnamefont {Patterson}},\ and\ \bibinfo
  {author} {\bibfnamefont {D.~G.}\ \bibnamefont {Green}},\ }\bibfield  {title}
  {\bibinfo {title} {Many-body theory calculations of positron scattering and
  annihilation in ${\mathrm{h}}_{2}$, ${\mathrm{n}}_{2}$, and
  ${\mathrm{ch}}_{4}$},\ }\href
  {https://doi.org/10.1103/PhysRevLett.130.263001} {\bibfield  {journal}
  {\bibinfo  {journal} {Phys. Rev. Lett.}\ }\textbf {\bibinfo {volume} {130}},\
  \bibinfo {pages} {263001} (\bibinfo {year} {2023})}\BibitemShut {NoStop}%
\bibitem [{\citenamefont {Cassidy}\ \emph
  {et~al.}(2024{\natexlab{b}})\citenamefont {Cassidy}, \citenamefont
  {Hofierka}, \citenamefont {Cunningham},\ and\ \citenamefont
  {Green}}]{Cassidy2024_2}%
  \BibitemOpen
  \bibfield  {author} {\bibinfo {author} {\bibfnamefont {J.~P.}\ \bibnamefont
  {Cassidy}}, \bibinfo {author} {\bibfnamefont {J.}~\bibnamefont {Hofierka}},
  \bibinfo {author} {\bibfnamefont {B.}~\bibnamefont {Cunningham}},\ and\
  \bibinfo {author} {\bibfnamefont {D.~G.}\ \bibnamefont {Green}},\ }\bibfield
  {title} {\bibinfo {title} {Many-body theory calculations of positronic-bonded
  molecular dianions},\ }\href {https://doi.org/10.1063/5.0188719} {\bibfield
  {journal} {\bibinfo  {journal} {The Journal of Chemical Physics}\ }\textbf
  {\bibinfo {volume} {160}},\ \bibinfo {pages} {084304} (\bibinfo {year}
  {2024}{\natexlab{b}})},\ \Eprint
  {https://arxiv.org/abs/https://pubs.aip.org/aip/jcp/article-pdf/doi/10.1063/5.0188719/19692847/084304\_1\_5.01887
  19.pdf}
  {https://pubs.aip.org/aip/jcp/article-pdf/doi/10.1063/5.0188719/19692847/084304\_1\_5.01887
  19.pdf} \BibitemShut {NoStop}%
\bibitem [{\citenamefont {Moreira}\ and\ \citenamefont
  {Bettega}(2024)}]{Moreira2024}%
  \BibitemOpen
  \bibfield  {author} {\bibinfo {author} {\bibfnamefont {G.}~\bibnamefont
  {Moreira}}\ and\ \bibinfo {author} {\bibfnamefont {M.}~\bibnamefont
  {Bettega}},\ }\bibfield  {title} {\bibinfo {title} {Can a positron bind to
  the para-benzoquinone molecule?},\ }\href
  {https://doi.org/10.1140/epjd/s10053-024-00800-2} {\bibfield  {journal}
  {\bibinfo  {journal} {Eur. Phys. J. D}\ }\textbf {\bibinfo {volume} {78}}
  (\bibinfo {year} {2024})}\BibitemShut {NoStop}%
\bibitem [{\citenamefont {Kendall}\ \emph {et~al.}(1992)\citenamefont
  {Kendall}, \citenamefont {Dunning~Jr},\ and\ \citenamefont
  {Harrison}}]{Dunning1992}%
  \BibitemOpen
  \bibfield  {author} {\bibinfo {author} {\bibfnamefont {R.~A.}\ \bibnamefont
  {Kendall}}, \bibinfo {author} {\bibfnamefont {T.}~\bibnamefont
  {Dunning~Jr}},\ and\ \bibinfo {author} {\bibfnamefont {R.~J.}\ \bibnamefont
  {Harrison}},\ }\bibfield  {title} {\bibinfo {title} {Electron affinities of
  the first-row atoms revisited. systematic basis sets and wave functions},\
  }\href@noop {} {\bibfield  {journal} {\bibinfo  {journal} {J. Chem. Phys.}\
  }\textbf {\bibinfo {volume} {96}},\ \bibinfo {pages} {6796} (\bibinfo {year}
  {1992})}\BibitemShut {NoStop}%
\bibitem [{NWC(2024)}]{NWChem}%
  \BibitemOpen
  \href {https://www.nwchem-sw.org/} {\bibinfo {title}
  {https://www.nwchem-sw.org/}} (\bibinfo {year} {2024})\BibitemShut {NoStop}%
\bibitem [{\citenamefont {Green}\ and\ \citenamefont
  {Gribakin}(2015)}]{Green2015}%
  \BibitemOpen
  \bibfield  {author} {\bibinfo {author} {\bibfnamefont {D.~G.}\ \bibnamefont
  {Green}}\ and\ \bibinfo {author} {\bibfnamefont {G.~F.}\ \bibnamefont
  {Gribakin}},\ }\bibfield  {title} {\bibinfo {title} {$\ensuremath{\gamma}$
  spectra and enhancement factors for positron annihilation with core
  electrons},\ }\href {https://doi.org/10.1103/PhysRevLett.114.093201}
  {\bibfield  {journal} {\bibinfo  {journal} {Phys. Rev. Lett.}\ }\textbf
  {\bibinfo {volume} {114}},\ \bibinfo {pages} {093201} (\bibinfo {year}
  {2015})}\BibitemShut {NoStop}%
\bibitem [{\citenamefont {Patterson}(2019)}]{Patterson2019}%
  \BibitemOpen
  \bibfield  {author} {\bibinfo {author} {\bibfnamefont {C.~H.}\ \bibnamefont
  {Patterson}},\ }\bibfield  {title} {\bibinfo {title} {Photoabsorption spectra
  of small {Na} clusters: {TDHF} and {BSE} versus {CI} and experiment},\
  }\href@noop {} {\bibfield  {journal} {\bibinfo  {journal} {Phys. Rev.
  Mater.}\ }\textbf {\bibinfo {volume} {3}},\ \bibinfo {pages} {043804}
  (\bibinfo {year} {2019})}\BibitemShut {NoStop}%
\bibitem [{\citenamefont {Patterson}(2020)}]{Patterson2020}%
  \BibitemOpen
  \bibfield  {author} {\bibinfo {author} {\bibfnamefont {C.~H.}\ \bibnamefont
  {Patterson}},\ }\bibfield  {title} {\bibinfo {title} {Density fitting in
  periodic systems: Application to tdhf in diamond and oxides},\ }\href
  {https://doi.org/10.1063/5.0014106} {\bibfield  {journal} {\bibinfo
  {journal} {The Journal of Chemical Physics}\ }\textbf {\bibinfo {volume}
  {153}},\ \bibinfo {pages} {064107} (\bibinfo {year} {2020})},\ \Eprint
  {https://arxiv.org/abs/https://pubs.aip.org/aip/jcp/article-pdf/doi/10.1063/5.0014106/15577721/064107\_1\_online.pdf}
  {https://pubs.aip.org/aip/jcp/article-pdf/doi/10.1063/5.0014106/15577721/064107\_1\_online.pdf}
  \BibitemShut {NoStop}%
\bibitem [{NIS(2024)}]{NIST}%
  \BibitemOpen
  \href {https://cccbdb.nist.gov/diplistx.asp} {\bibinfo {title}
  {https://cccbdb.nist.gov/diplistx.asp}} (\bibinfo {year} {2024})\BibitemShut
  {NoStop}%
\bibitem [{\citenamefont {Lide}(2005)}]{CRC86}%
  \BibitemOpen
  \bibinfo {editor} {\bibfnamefont {D.~R.}\ \bibnamefont {Lide}},\ ed.,\
  \href@noop {} {\emph {\bibinfo {title} {CRC Handbook of Chemistry and
  Physics, 97th ed.}}}\ (\bibinfo  {publisher} {CRC Press},\ \bibinfo {year}
  {2005})\BibitemShut {NoStop}%
\bibitem [{\citenamefont {Dzuba}\ and\ \citenamefont
  {Gribakin}(1994)}]{Dzuba1994}%
  \BibitemOpen
  \bibfield  {author} {\bibinfo {author} {\bibfnamefont {V.~A.}\ \bibnamefont
  {Dzuba}}\ and\ \bibinfo {author} {\bibfnamefont {G.~F.}\ \bibnamefont
  {Gribakin}},\ }\bibfield  {title} {\bibinfo {title} {Correlation-potential
  method for negative ions and electron scattering},\ }\href
  {https://doi.org/10.1103/PhysRevA.49.2483} {\bibfield  {journal} {\bibinfo
  {journal} {Phys. Rev. A}\ }\textbf {\bibinfo {volume} {49}},\ \bibinfo
  {pages} {2483} (\bibinfo {year} {1994})}\BibitemShut {NoStop}%
\end{thebibliography}

%

\onecolumngrid

\vspace{20pt}

\end{document}